\documentclass{amsart}
\usepackage{amsmath,amsfonts,tikz,bm,hyperref,graphicx}
\renewcommand\vec[1]{\bm{\mathrm{#1}}}
\newtheorem{define}{Definition}[section]

\title{Token Exchange Games}
\author{Viroshan Naicker}
\email{viroshan@northquotient.com}
\date{\today}

\begin{document}

\begin{abstract}
	Human societies engage in a number of games which use tokens as a means to allocate, issue and access gated resources and property rights:~The notion of exchanging tokens that represent and carry value from the past into the future to facilitate economic exchange is a fundamental concept. As agents exchange tokens a network structure is created in which the tokens move from agent to agent. Agents form the vertices in the network and the exchange of tokens creates the link structure. The rules of exchange adopted collectively by agents shape the network structure and affect the distribution of tokens amongst agents. Many inventions including banking, payments networks and blockchain technology have been created for the purpose of keeping accurate records of these implicit networks. However, our formal understanding of tokenised systems as large scale, iterative, interacting structures is limited. The aim of this paper is to study the dynamics of token exchanges as a game, and introduce a mathematical framework for understanding ledgers as complex networks.
\end{abstract}


\maketitle	

\section{Introduction}

Tokens that carry information play a fundamental role in how societies allocate, issue and access gated resources and property rights:~Tokens act as records which facilitate the rapid exchange of rights from agent to agent in an economy. Further, tokens carry value from the past to the future and create complex large scale interacting systems:~As tokens move between economic agents, an evolving network structure is created in which agents form the nodes of the network and the pairwise exchange of tokens form the edges of the network. 
Ledgers, in all their forms:~from ancient tax logbooks to blockchains, have been used throughout history to keep track of the distribution of tokens in these types of network games. Moreover, our  information rich, connected, world offers opportunities for near instant exchange across multiple token types. However, despite our widespread use of tokens, our formal understanding of tokenized systems as large scale, iterative, interacting networks is limited. 

This problem may be posed abstractly: Given a set of identical tokens that are allocated to agents and a system of rules for how these tokens are allowed to move between agents, how will the rules influence the distribution of tokens after many rounds of exchanges? And, given multiple, distinct, sets of tokens that are allocated to agents and a system of rules for determining exchanges between these distinct sets of tokens, how will the rules influence the distribution of tokens across different token types after many rounds of exchanges?  

The aim of this paper is to answer the above questions in a mathematical context.
In order to do so, we use graph theory and linear algebra \cite{Bollobas:1986:CSS:7228,Nagy:2012,West:2017,Wilson:2015} to define a token exchange game as family of networks that evolve according to a discrete parameter (Section~\ref{teg_ledger}). As the network evolves the token distribution, a ledger, is updated accordingly. In order to apply the theory, we analyse games involving single token types (Section~\ref{slg}) and show that the celebrated PageRank algorithm \cite{langville2011google} can be viewed as a token exchange game. Further, using information theory \cite{cover2012elements}, we consider the impact of network behaviour on token distributions (Section~\ref{td}). Since most games involve the exchange of tokens of different types amongst agents and the notion of fungibility, the behaviour of exchanges across multiple networks is considered in depth (Section~\ref{mlg}). We develop several natural metrics related to token velocity (Section~\ref{mm}). Finally, we construct examples of token exchange games that are drawn from blockchain technology (Section~\ref{ex}). In particular, we consider the Lightning Network \cite{LNW:2016} and the Circles token system \cite{CRC}. 


\section{Notation}

Various pieces of notation from set theory, graph theory and linear algebra are used throughout the paper. For clarity sake, a few of the conventions used are highlighted below. 

\subsection{Set Theory}

For set theory and combinatorics, we largely follow the notation set out in \cite{Bollobas:1986:CSS:7228}. The following labelling conventions hold for sets: Discrete sets of elements are labelled using upper case Latin letters, while elements of sets are labelled using lower case Latin letters. Families of sets (set-systems) are labelled using calligraphic upper case letters. An index set will mean a set  $I=\{0,1,2,\ldots n\}$ which serves as a parameter space; if we say $r\in I$ it simply means that the parameter $r$ ranges over the values in the index $I$. 

\subsection{Metric Spaces}

The formal concept of distance is key to understanding a ledger as an abstract object, so language from metric spaces will be necessary. A detailed introduction may be found in \cite{Apostol:1974,Rudin:1974}.
For our purposes a metric space $(M,d)$ is a topological space $M$ equipped with an additional notion of distance $d$ between elements of $M$. Given elements $x$, $y$, $z$ $\in$ $M$, $d(x,y)$ is the distance between $x$ and $y$ such that $d(x,y)\ge 0$ with equality if and only if $x=y$, $d(x,y)=d(y,x)$, and $d(x,z)\le d(x,y)+d(y,z)$. 

A sequence $k=k_{1}, k_{2}, k_{3}\ldots $ will be denoted $k=(k_{n})$ where $n$ is a finite or countably infinite parameter. Similarly, an array will be denoted $[k_{nm}]$ where $n$ and $m$ are finite or countably infinite parameters. 

\subsection{Linear Algebra}

Most of this paper draws on basic linear algebra and the notation follows, to an extent, the notes \cite{Nagy:2012}. Vectors will usually be written as bold font Latin letters and matrices as upper case Latin letters, for example, ${\bf x}$ is a vector and $A$ is a matrix. If the components of a vector or matrix are emphasised we will write ${\bf x}=[x_{i}]=(x_{1},x_{2},\ldots,x_{m})^{T}$ or $A=[A_{ij}]$ where the dimensionality depends on the underlying space and the superscript $T$ denotes the transpose. In order to keep track of several related matrices and vectors, a subscript notation will be used, for example, ${\bf x}_{r}\in {\mathbb R}^{m}$ refers a vector with $m$ components parametrized by $r\in I$ where $I$ is an index set. Similarly $A_{rs}$ is a doubly parametrized collection of matrices with $r\in I$ and $s\in J$; and each matrix is written in components as $A_{rs}=[A_{ij}]_{rs}$. Effectively, this is a tensor with four indices but notationally it is more useful in the following to think of it as a matrix with two parameters. A \emph{left stochastic matrix} is a matrix whose columns sum to one \cite{Norris:1998}. 

\subsection{Graph Theory}

Notation and definitions in graph theory tend to vary quite widely, so as a baseline we will define necessary notation in this section. General introductions to graph theory may be found in \cite{West:2017, Wilson:2015}. More advanced material on random graphs and networks may be found in \cite{Bollobas:2009,bornholdt2006handbook,Chung_graphtheory}.

A \emph{graph} $G$ is a (usually) finite of \emph{vertices} $V(G)$ and a set of \emph{edges} $E(G)$ which comprise pairs of vertices. We may write $V(G)=\{v_{1}, v_{2}, \ldots, v_{n}\}$ and so the vertex $v_{i}\in V(G)$. Similarly, $e=v_{i}v_{j}\in E(G)$ is the edge between vertices $v_{i}$ and $v_{j}$. The total number of vertices, $|V(G)|$, is known as the \emph{order} of $G$, and the total number of edges, $|E(G)|$, is known as the \emph{size} of $G$. 

There are several standard classes of graphs. An undirected graph that has no loops or multiple edges is called a \emph{simple graph} or just a graph. A graph with directed edges such that $v_{i}v_{j}\ne v_{j}v_{i}$ is called a \emph{directed graph}. A graph with directed edges that allows \emph{self-loops} (edges such as $v_{i}v_{i}$) is called a \emph{pseudo-graph}. In a directed graph, if the edge $v_{i}v_{j}$ starts at $v_{i}$, then $v_{i}$ is called the head, and $v_{j}$ is called the tail. Largely, we will work with pseudo-graphs and for brevity sake we will refer to them as graphs. In the event that we require simple or directed graphs this will be explicitly stated.

Vertices $v_{i}$ and $v_{j}$ are \emph{adjacent} if $v_{i}v_{j}\in E(G)$. In a simple graph, the \emph{open neighbourhood} of a vertex $v_{i}$ is the set of vertices $N(v_{i})$ such that for $v_{j}\in N(v_{i})$ the edge $v_{i}v_{j}\in E(G)$. The \emph{closed neighbourhood} of $v_{i}$ is the set $N(v_{i})\cup\{v_{i}\}$ and is denoted $N[v_{i}]$. 

In a directed or pseudo-graph, given any vertex $v_{i}\in V(G)$, the \emph{open out-neighbourhood} of $v_{i}$ is denoted $N(v_{i})$ and consists of vertices $v_{j}$ such that $v_{i}v_{j}\in E(G)$ and the edge $v_{i}v_{j}$ starts at $v_{i}$ and ends at $v_{j}$ and $v_i\not\in N(v_{i})$. The \emph{closed out-neighbourhood} of $v_{i}$ is $N[v_{i}]$ is defined similarly but $v_{i}\in N[v_{i}]$. Notice that we can partition the edge set $E(G)$ of a directed graph and a pseudo-directed graph using out-neighbourhoods. 

The \emph{degree} of a vertex is the number of edges that are incident with a vertex. In a simple graph we refer only to the degree of the vertex. In a directed graph the \emph{out-degree} of a vertex counts the number of edges for which the vertex is the head, and the in-degree of a vertex counts the number of edges for which the vertex is the tail. In a pseudo-graph, a loop counts as both part of the out-degree of the vertex and the in-degree of a vertex. 

A \emph{tree} is a simple or directed graph which contains no cycles.
A \emph{bipartite graph} is a simple or directed graph which has partite sets $X$ and $Y$, so that if both $v_{i}$ and $v_{j}\in X$ or both $v_{i}$ and $v_{j}\in Y$ then $v_{i}v_{j}, v_{j}v_{i}\notin E(G)$.


\section{Token Exchange Games}
\label{teg_ledger}

\subsection{Tokens}

In order to introduce the idea of a token exchange game, we need to be specific about what we will mean by a token and a token system. Generally, there is a tendency to define tokens by their properties, so, for example, monetary tokens are required to be durable, divisible, transferrable and scarce. Alternatively, monetary tokens may be defined in terms of their fungibility with goods and services: Money is exchangeable for `real objects'. A general conceptual framework is to think of a token as an \emph{allocatable property right} -- a token may be allocated to an agent participating in its native token system and, concomitantly, the rights that are imbued to the token by the system which are active may be exercised by the agent who owns the token. Within this framework, we motivate the next definitions. 

\begin{define}
	Token. A token is an information record that may be stored in an arbitrary medium. Further, identical tokens may be stored on different media. Tokens may be allocated to agents and if the information contained in the token allocates a right to the agent, then the agent is said to be the owner of that right and may exercise that right. Different token-types allocate different rights to agents and are thus distinguishable. 
\end{define}

\begin{define}
	Token System. A token system is collection of identical tokens and a rule set for governing token behaviour and distribution properties amongst agents participating in the system. Further, different token-types belong to distinguishable token systems. 
\end{define}

In the token systems that we will examine, a basic premise is that agents have the right of exclusive ownership and, as such, multiple agents are not allowed to own the same token at the same time. Also, the tokens that we will deal with will, in general, be transferrable from agent to agent. Non-transferrable tokens do arise in practice and these are a special case of the model. Notice that with the definitions above, an apple is as much a token as a coin used to pay for that apple. The difference is that the information contained in a tokenised apple will degrade after some period and coins are usually recorded on more durable media. 

\subsection{Ledgers} Since agents may own and transfer tokens and tokens may be imbued with valuable rights, the record keeping of the distribution of tokens amongst agents in a given token system is of paramount importance. Ledgers are the means of keeping track. There are two intuitive ways to articulate the idea of a ledger: Firstly, a ledger can be thought of as the token system itself simply because a set of tokens distributed amongst a collection of agents is a record of the token distribution; each agent owns some tokens and collectively they make up the distribution. Alternatively, a ledger can be thought of as a snapshot of the token distribution at a given point in time that is a separate information record from the tokens allocated to agents; so a secondary record. In either case, since we take it as a given that agents can own (and be allocated) their tokens exclusively, a ledger for us is a set of labels with a magnitude associated to each label. More specifically, we are interested in ledgers that evolve in time:~Sequences of ledgers that can be ordered, and that track changes to the token distribution due to the behaviour of participating agent as they make token exchanges within the constraints of the token system. The definitions below formalise the notion of a ledger by combining a set of labels with a surjective map to a metric space.

\begin{define}
	\label{rank}
	Ranking. Let ${S}$ be a set of labels and $(M,d)$ a metric space. If for every $s\in {S}$ there is a surjective mapping $f: s\rightarrow f(s)\in M$ and, for every pair $s,t\in S$ and a metric $d$ on $M$ the distance $d(f(s),f(t))<\infty$, then the triple $\{{S },f,(M,d)\}$ defines a ranking. More specifically, since $f$ and $d$ define a pairwise distance between elements of ${S}$ they may be used to order the members of $S$.
\end{define} 

\noindent Next, the idea of an ranking associated to a set of labels allows the formal definition of a ledger as a collection of pairs.  

\begin{define}
	Ledger. Suppose that $\{{S},f,(M,d)\}$ is a ranking. A ledger ${L}$ is a collection of pairs such that  $\{s,f(s)\}\in {L}$ for $s\in {S}$.
\end{define}

\noindent In practice, most ledgers use $M={\mathbb R}$ as the base metric space with the metrics $d(x,y)=\sqrt{x^{2}+y^{2}}$ or $d(x,y)=|x-y|$ for $x,y\in {\mathbb R}$ to give a notion of distance.
\noindent In addition, ledgers may be accompanied by a finite set of physical tokens which represent the ledger and approximates the value recorded in the ledger. Physical money, for example, is grainy and prices adjust to this graininess. This motivates the definitions of a \emph{token set} and a \emph{tokenised ledger}. 
\begin{define}	
	Token Set. Let $T=\{t_{1},t_{2},\ldots, t_{k}\}$ be a set of real numbers such that $t_{i}<t_{j}$ for $i<j$ and suppose that  $\{{ S},f,(\mathbb{R},d)\}$ is a real valued ranking. Suppose for every $s\in S$, coefficients $m_{i}\in {\mathbb N}$, and a fixed $\epsilon$ that there exists an approximation of $f(s)$ in terms of $T$ as $f(s)\approx \sum_{i} m_{i}t_{i}$ such that $|f(s)-\sum_{i}m_{i}{t_{i}|\le t_{1}\le \epsilon}$. Then, $T$ is a valid token set for the ranking $\{{S},f,(\mathbb{R},d)\}$.
\end{define}
\begin{define}
	Tokenised Ledger. Suppose that $\{{S},f,({\mathbb R},d)\}$ defines a ranking and that $L$ is the corresponding ledger. Let $T$ be a valid token set for the ranking $\{{S},f,({\mathbb R},d)\}$. Then, a representation of each $f(s)$  in ${L}$ using the members of $T$ is called a tokenisation of ${L}$ and ${L}$ is a tokenised ledger with respect to the set $T$.
\end{define}
A further point worth noting is that when ledgers are tokenised, then entries in the ledger are often rounded in order to take into account the graininess of the underlying token set. In this case, it is unnecessary to distinguish between a ledger and its tokenisation. This is the approach that we will take for the rest of the paper. 
\begin{define}
	\label{seq}
	Ledger Sequence. Suppose that $t \in I$, where $I$ is an index set. A ledger sequence is a sequence of ledgers in the parameter $t$ such that for each $t$ there exists a ledger $L_{t}$. Moreover, the ledgers are ordered in a sequence $(L_{t})=(L_{0}, L_{1}, L_{2}\ldots)$. If $I$ is a finite set, then $(L_{t})$ terminates for some finite $k$. 
\end{define}

\noindent Using Definitions~(\ref{rank})-(\ref{seq}) a monetary system of coins and notes, for example, is a tokenised ledger in which the token set is made up of the coins and notes. An agent's balance in the ledger is represented by these denominated coins and notes and the information record is a physical one in which agents are assigned their money as a property right. 
Socially, ledgers act as a shared information record and the tokenisation of ledgers allows for agents to rapidly alter the distribution of values across multiple ledgers by means of exchange.
In monetary economies property rights are allocated to the tokens in a given `currency' ledger, and this ledger serves to facilitate the reallocation of property rights amongst economic agents. 
This redistribution process, and the movement of tokens, induces ledger sequences as maps in which $L_{i}\mapsto L_{i+1}$ according to the `rules' by which tokens are allowed to be exchanged amongst token owners. 
This type of dynamic behaviour within a formal ledger motivates the idea of a \emph{token exchange game}:~a game in which agents alter the distribution of values in a ledger by exchanging tokens amongst themselves according to the rules of the token system.
In the next sections we develop such a model using the languages of graph theory and linear algebra.  

\subsection{Network Representation}

A natural context for modelling ledgers is as a record of exchanges in a network. In this section, we develop notation for describing a family of dynamic ledgers as an evolving family of pseudo-graphs (directed with loops allowed). For generality sake we introduce a two parameter family of graphs:~The first parameter $r$ keeps track of round by round ledger evolution and the second parameter $s$ keeps track of different ledger labels. This is to allow us to consider evolving ledgers of different token types and systems with different internal rule sets that are interacting dynamically with each other. 

\begin{define}
	Two Parameter Graph Family. Let ${\mathcal G}(r,s)$ be a two parameter family of graphs such that $r\in I$ is a countably infinite or finite index set and $s\in J$ is a set of labels. A member graph of the family is written as $G_{rs}=(V_{rs},E_{rs})$ which has vertex set $V_{rs}$ and edge set $E_{rs}$. Furthermore, set $|V_{rs}|=n_{rs}$ and $|E_{rs}|=m_{rs}$. 
	\label{dpg}
\end{define}

\begin{define}
	Vertex Weighting. Let $G=(V,E)$ be a graph with vertex set $V$ and edge set $E$. Given any subset $S\subseteq V$ a vertex weighting of $S$ is a map $\phi: S\rightarrow \mathbb{R}_{\ge 0}$, such that for $v\in S$, $\phi(v)\in [0,\infty)$ and for any subset $S'\subseteq S$ we have $\phi(S')=\sum_{v\in S'}\phi(v)$. Thus, $\phi(V)=\sum_{v\in V}\phi(v)$ and, similarly, $\phi(S)=\sum_{v\in S}\phi(v)$ for any subset $S\subseteq V$. A vertex weighting on $S\subseteq V$ is non-trivial if and only if $\phi(S)> 0$.
	\label{vw}
\end{define}
\noindent Note that the vertex weightings we use below will always be non-trivial. 
\begin{define}
	Edge Weighting. Let $G=(V,E)$ be a graph with vertex set $V$ and edge set $E$ with $|V|=n$. Suppose that $S\subseteq V$. Let $F(S,N[S])$ be the set of edges that have $u\in S$ as the head, and $v\in N[S]$ as the tail. Suppose $f=uv\in F(S,N[S])$. An edge weighting on $F(S, N[S])$ is a map $\psi(S,N[S])$ such that the following conditions hold:
	\begin{itemize}
		\item[$(i)$] For $f\in F$, $0< \psi(u,v)\le 1$ and $\psi(u,v)=0$ iff $f\notin F$.  
		\item[$(ii)$]  $\sum_{f\in F(S,N[S])} \psi(u,v)=1$.
		\item[$(iii)$] If $S=\{u\}$ and $v\in N[u]$, then $\sum_{u\in V, v \in N[u]}\psi(u,v)=n$.
	\end{itemize}
	\label{ew}
\end{define}

Note that it is easy to show that in Defintion~(\ref{ew}) conditions $(i)$ and $(ii)$ imply condition $(iii)$, since each vertex and its closed out-neighbourhood have weight one, and there are $n$ vertices or, alternatively, each $e\in E$ belongs to a unique $N[u]$ for some $u\in V$, and we sum the weights over $n$ closed out-neighbourhoods. We have stated it this way, since we will mainly be interested in edge weightings of vertex neighbourhoods over the entire vertex set $V$.  

\subsection{Algebraic Representation} Definitions~(\ref{vw}) and (\ref{ew}) above are readily adapted to a linear algebraic representation. In order to move from a network picture to a linear algebraic picture we need consider the vertices of the graph as positions that are assigned weights and the edges as matrix entries with their edge weightings as magnitudes. This motivates the next definitions of a vertex weighting vector and an edge weighting matrix.  

\begin{define}
	Vertex Weighting Vector.  Let $G=(V,E)$ be a graph and suppose that $V=\{v_{1},v_{2},\ldots,v_{n}\}$ is a labelling of the vertex set. Let $\phi$ be a vertex weighting on $V$. The vertex weighting vector that corresponds to $(G,\phi)$ is given by $\vec{x}=(x_{1},x_{2},\ldots, x_{n})^{T}$ where $x_{i}=\phi(v_{i})$ for $i=1,2,\ldots n$. 
	\label{vv}
\end{define}

\begin{define}
	Edge Weighting Matrix. Let $G=(V,E)$ be a graph and suppose that $V=\{v_{1},v_{2},\ldots,v_{n}\}$ is a labelling of the vertex set. Suppose that $e_{ij}=v_{i}v_{j}$  where $e_{ij}\in E$. Let $\psi$ be an edge weighting on $G$ and let $\psi(e_{ij})=w_{ij}$. Then, $0< w_{ij}\le 1$ if $e_{ij}\in E$ and $w_{ij}=0$ for $e_{ij}\notin E$. Setting $W=[w_{ij}]$ gives the edge weighting matrix $W$ which has dimension $n\times n$ and is left stochastic ($\sum_{j}w_{ij}=1$) due to the properties of edge weightings. 
	\label{em}
\end{define}

\subsection{Token Exchange Games} Using the definitions above, we are now in the position to present a \emph{token exchange game} as an abstract model of a ledger in which a finite number of tokens (possibly of different types) are distributed and exchanged amongst a collection of players. We are interested in describing a ledger sequence $(L_{r})$ where $r\in I$. The sequence keeps track of the changes in token distribution amongst a set of players and where each player can make a limited `hold token' or `transfer token' decision in a given round of the game. 
This sets up a framework for describing an abstract class of games in which tokens move from agent to agent according to an arbitrary rule set without the possibility of double spending. Moreover, the framework can be extended to consider the behaviour of multiple interacting ledgers by introducing a second parameter $s\in J$. 

\begin{define} 
	\label{teg}
	Token Exchange Game. Let ${\mathcal G}(r,s)$ be a two parameter graph family such that for each member graph $G_{rs}=(V_{rs}, E_{rs})$ a valid vertex weighting $\phi_{rs}$ and a valid edge weighting $\psi_{rs}$ exists for each $v\in V_{rs}$ and $e\in E_{rs}$. Given an initial vertex weighting  $\phi_{0s}$ for $r=0$ and for each $s$, a token exchange game is an iterative game where for $u\in V_{rs}$ and $v\in N[u]$:
	\begin{equation}\phi_{(r+1)s}(u)=\sum_{u\in N[v]}\psi_{rs}(v,u)\phi_{rs}(v).\label{tegeq1}\end{equation} Alternatively, given a labelling of $V_{rs}$, a vertex weighting vector $\vec{x}_{rs}$ and an edge weighting matrix $W_{rs}=[w_{ij}]_{rs}$, a token exchange game is defined iteratively using matrix multiplication as: 
	\begin{equation}\vec{x}_{(r+1)s}=W_{rs}\vec{x}_{rs},\label{tegeq2}\end{equation}  where the initial vertex weighting vector, $\vec{x}_{0s}$, is known at $r=0$.

\end{define}

The equivalence of the network and matrix representations of a token exchange game follows directly from the definition of matrix multiplication. In the network picture, the focus is at vertex and subset level while the matrix picture gives a view of the dynamics of the system as a whole. The rules of the token exchange game determine the nature of $W_{rs}$ and $\psi_{rs}$. Conditions such transaction costs and token issuance mechanisms can be expressed as mathematical rules contained in $W_{rs}$ and $\psi_{rs}$ that depend on $r$ and alter the distribution of tokens iteratively through~Defintion~(\ref{teg}). The next definition explains the parameter $s$ in a token exchange game.
\begin{define}
	Layers. Given a token exchange game each $s\in J$ defines a layer the game. We make the following distinctions between layers:
	\begin{itemize}
		\item[(i)] A single layer token exchange game is a game in which $|J| = 1$ and
		there is only one available token type in the game. 
		\item[(ii)] A multilayer token exchange game allows for $|J|$ concurrent layers each
		of which represents a unique token type. 
		\item[(iii)] A sublayer of a token exchange game  uses a two parameter graph subfamily, for example, $\mathcal{F}(r,s)\subset \mathcal{G}(r,s)$ where $r\in I^{*}\subseteq I$ and $s\in J^{*}\subseteq J$, to construct a nested token exchange game. 
	\end{itemize}
\end{define}
Finally, we distinguish between games that are open and games that are closed using the token supply. It is easy to see that in a given round $r$, for any given $s$, the \emph{token supply} (of type-$s$ tokens) is: \begin{equation}|\vec{x}_{rs}|=\sum_{i}x_{i}=\sum_{v\in V_{rs}}\phi_{rs}(v),\end{equation} using Definitions~(\ref{vw}), (\ref{vv}) and (\ref{teg}). The next question is whether tokens can be introduced into the game from an external supply or taken out of the game. Thus, we make the distinction below. 
\begin{define}
	\label{oc}
	Open and Closed Games. For a fixed $s$, let $T_{rs}=|\vec{x_{rs}}|$ be the token supply for a token exchange game defined according to (\ref{teg}). If $T_{rs}=T_{s}$ and remains constant for all $r$, then the game is called closed, otherwise the game is called open. 
\end{define}
In essence an open game allows for changes to the token supply during the game. One way to get around using an open game is to assume that all the tokens that will ever be created are held in a special treasury wallet and released to (or locked away from) the `rest of the world' for the appropriate round.  However, it may be useful to study an open game as a subset of a larger closed game, as well as allowing tokens to be minted and destroyed. In order to model an open game another class of vertex weightings are required, and these are constructed so that no player may have a negative balance even after tokens are deducted by an external party. 
\begin{define}
	\label{rvw}
	Relative Vertex Weighting. Let $G=(V,E)$ be a graph and suppose that $\phi$ is a vertex weighting on $V$ and $\psi$ is an edge weighting on $E$. A relative vertex weighting $\pi$ is a vertex weighing on $V$ with the positivity condition relaxed so that  $\pi(v)\in [-\phi(v)\times \psi(v,v),\infty)$ for any $v\in V$.  
\end{define}
\begin{define}
	Relative Vertex Weighting Vector. Let $G=(V,E)$ be a graph and suppose that $\pi$ is a relative vertex weighting on $V$. Let $V=\{v_{1},v_{2},\ldots, v_{n}\}$. The relative vertex weighting vector corresponding to  $(G,\pi)$ is given by $\vec{y}=(y_{1},y_{2},\ldots,y_{n})^{T}$ where $\pi(v_{i})=y_{i}$ and $i=1,2,\ldots, n$. 
\end{define}
\begin{define}
	\label{oteg}
	Open Token Exchange Game. Let ${\mathcal G}(r,s)$ be a two parameter graph family such that for each member graph $G_{rs}=(V_{rs}, E_{rs})$ a valid vertex weighting $\phi_{rs}$ and a valid edge weighting $\psi_{rs}$ exists for each $v\in V_{rs}$ and $e\in E_{rs}$. An open token exchange game is a token exchange game with a relative vertex weighting $\pi_{rs}$ on $G_{rs}$ where for $u\in V_{rs}$ and $v\in N[u]$ \begin{equation}\phi_{(r+1)s}(u)=\pi_{rs}(u)+\sum_{u\in N[v]}\psi_{rs}(v,u)\phi_{rs}(v).\end{equation}Alternatively, given a labelling of $V_{rs}$, a vertex weighting vector $\vec{x}_{rs}$,  a relative vertex weighting vector $\vec{y}_{rs}$ on $G_{rs}$,  and an edge weighting matrix $W_{rs}=[w_{ij}]_{rs}$ an open token exchange game is defined iteratively using matrix multiplication as: 
	\begin{equation}\vec{x}_{(r+1)s}=W_{rs}\vec{x}_{rs}+\vec{y}_{rs}.\label{otegeq2}\end{equation}Further, in each case, the respective initial states $\phi_{0s}$, $\pi_{0}$ and $x_{0s}$, $y_{0s}$ are known when $r=0$.

\end{define}
In an open token exchange game the additional vertex weighting allows for the creation and destruction of tokens through an external source. This can be useful, for example, when considering token airdrop strategies. For brevity sake, in the rest of the paper we will refer to a token exchange game as simply a game, an open game or closed game, as appropriate. 


\section{Single Layer Games}
\label{slg}

A single layer game may be modelled by a one parameter family of graphs ${\mathcal G}(r)$ where $r\in I$, and $I=\{0,1,2,\ldots k\}$. This represents a game with $k+1$ rounds. Primarily, we will make use of the matrix descriptions given by Eqs. (\ref{tegeq2}) and (\ref{otegeq2}). We rewrite them as:
\begin{equation}
\label{nes}\vec{x}_{r+1}=W_{r}\vec{x}_{r}
\end{equation}
and
\begin{equation}
\label{es}\vec{x}_{r+1}=W_{r}\vec{x}_{r}+\vec{y}_{r}.
\end{equation}
Ranging over the index $r\in I$ gives us $k+1$ matrices $W_{r}$ and $k+1$ vectors $\vec{x}_{r}$ and $k+1$ vectors $\vec{y}_{r}$.~Each graph $G_{r}$ has order $n_{r}$ and size $m_{r}$. This fixes the dimensionality of the matrices and vectors above as $n_{r}\times n_{r}$ and $n_{r}\times 1$ respectively. Further, the matrix $W_{r}$ must have $m_{r}$ entries. We next consider two examples of single layer token exchange games. 

\subsection{PageRank}

There are direct similarities between Eq.~(\ref{es}) and the construction of the PageRank algorithm used by Google \cite{langville2011google}. In the PageRank model for the internet, webpages are the vertices of a directed graph and the (possibly multiple) links between pages are directed edges. The main assumption being that webpages that are more important will have more links pointing to them and, therefore, a high probability of attracting traffic. The algorithm outputs a score for each webpage that depends on the chance a random surfer will arrive there. 

In order to adapt a simplified version of PageRank to a token based model, suppose that links between websites are transactions and that tokens are surfers. Each round a surfer either moves to another page or goes offline. Let ${\mathcal G}(r)$ where $r\in I$ be the family of graphs constructed in this way, and suppose that the graphs $G_{r}=(V_{r}, E_{r})$ have order $n_{r}$ and size $m_{r}$. Thus, there are $n_{r}$ webpages and $m_{r}$ links in round $r$. Let each webpage that exists in round $r$ be represented by a vertex in $V_{r}$. Suppose $v_{i}$, $v_{j}\in V_{r}$. Let $\ell (v_{i},v_{j})$ count the number of links from $v_{i}$ to $v_{j}$. By design, $\ell(v_{i},v_{i})=0$. If webpage $v_{i}$ has at least one link to webpage $v_{j}$ and $\ell(v_{i},v_{j})>0$, then the directed edge $v_{i}v_{j}\in E_{r}$. Moreover, let $k_{i}=deg(v_{i})$ be the total number of outbound links from the website represented by $v_{i}$. Set the edge weightings $\psi_{r}(v_{i},v_{j})=\ell(v_{i},v_{j})/k_{i}$. The initial vertex weighting may be constructed as $\phi_{0}(v_{i})=1/n_{0}$. Thus, there is exactly one token and this token is distributed equally amongst all the players. 

Now, if there is no possibility of a surfer leaving, then the game is closed and Eq.~(\ref{es}) may be used to iteratively calculate the PageRank.
In order to allow for surfers to enter and leave the game, suppose that a surfer leaves with probability $1-p$ (stays with probability $p$), the initial weighting is rescaled to $p\phi_{0}(v_{i})=p/n_{0}$ and a relative vertex weighting $\pi_{r}(v_{i})=(1-p)/n_{r}$ is added to each node in order to add a fractional surfer distributed evenly across the players. Direct construction of the vertex weighting vectors $\vec{x}_{r}$ and $\vec{y}_{r}$, and the matrix $W_{r}$ gives the PageRank iteratively via Eq.~(\ref{es}). The factor $p$ is commonly known as the damping factor \cite{BRIN1998107}. For further details on PageRank see \cite{BRIN1998107,DBLP,Richardson:2002:MKS:775047.775057}.

\subsection{Universal Basic Incomes}

Universal Basic Income systems have recently become popular amongst academics and governments \cite{standing2017basic,van2017basic,UBU:2017wp}. Consider the simple system where a treasury issues tokens to players and issued tokens dissipate from players wallets back to the treasury at a fixed rate as a counter inflationary measure. The dissipation effect ensures scarcity in the token and at the point where the dissipation rate is equal to the issue rate there is an equilibrium. 
The two player game below illustrates this idea as a token exchange game: One player represents the `treasury node' and the other player represents `the rest of the system'. Since the interactions between the treasury and the rest of the system are deterministic an explicit solution can be found for the player's token balances. 

Suppose that the players are called $A$ and $B$. Each round player $A$ gives a fixed number of tokens $f$ to player $B$. Player $B$ return a proportion of his total tokens $\epsilon$ to player $A$ each round of the game. Suppose that the game has the initial vector $\vec{x}_{0}=(x_{0}^{A},x_{0}^{B})^{T}$ and continues for $k+1$ rounds. Let $f=\delta x_{0}^{A}$ where $0\le \epsilon \le 1$ and  $0 \le \delta\le 1$. The matrices $W_{0}$, $W_{2}$, \ldots $W_{k+1}$ that govern the game can be obtained from the recurrence relationship
\[W_{j}=
\begin{bmatrix}
1-f/x^{A}_{j}& \epsilon\\	
f/x^{A}_{j} & 1-\epsilon
\end{bmatrix}\cdot
\]
The general token balances of each player are given by the recurrence equations
\[
\begin{matrix}
x^{A}_{j+1}&=&x_{j}^{A}+\epsilon x^{B}_{j}-f,\\
x^{B}_{j+1}&=&(1-\epsilon)x^{B}_{j}+f.
\end{matrix}
\]
The token balance of player $B$ at round $j$ can be solved as 
\[x^{B}_{j}= (1-\epsilon)^{j}x^{B}_{0}+f\left(\frac{1-(1-\epsilon)^{j}}{\epsilon}\right)\cdot
\]
If we suppose that $x_{0}^{B}=0$ and $x_{0}^{A}=\Omega$, then the the balance for player $A$ at round $j$ is given as
\[x_{j}^{A}=\Omega-f\times\left(\frac{1-(1-\epsilon)^{j}}{\epsilon}\right)
\]
Of course, a more complex system will have more than two players, but this is instructive: Player $B$ could be considered as an open game sublayer nested within this larger closed game. In the sublayer,  players may exchange tokens amongst themselves, but the core relationship between the treasury and the rest is captured here. 

\subsection{Remarks} 

Single layer games provide a useful construction for understanding payment systems and networks as discretely parametrised evolving graphs. The direct use case for this type of language is monetary systems including cryptocurrencies and cryptotoken systems: These include altcoins, utility coins, security tokens and token governance models. In the case of blockchain based ledgers each block of transactions could be considered a round in a token exchange game with wallet addresses representing agents or vertices. As tokens are mined the system may be treated as open using Eq.~(\ref{es}) with a random `miner node' receiving a token reward per round. Alternatively, an artificial treasury node may be introduced and the system may be considered closed with the treasury holding all the tokens that may ever be put into circulation. 
In terms of empirical work, public blockchain ledgers like Bitcoin \cite{Nakamoto_bitcoin:a} will allow the construction of historical exchange matrices from blockchain data. 


\section{Token Distributions}
\label{td}

Using a vector to represent the tokens owned by players has an interesting probabilistic interpretation:~Suppose that {\bf v} is a vector which represents a token allocation amongst $n$ players. Thus, $\vec{v}\in\mathbb{R}^{n}$ and ${\bf v}=[v_{i}]$ in components where $i=1,\ldots,n$. The total number of tokens in circulation is given by $|{\bf v}|=\sum_{i=1}^{n} v_{i}$. As a result we can normalize the token allocation ${\bf v}$ and create a token distribution. Setting
\begin{equation}
\label{tdist}
p_{i}=\frac{v_{i}}{|{\bf v}|}
\end{equation}
gives the probability distribution $p(i)=p_{i}$. This may be interpreted as follows:~Suppose that a token is drawn randomly from a bin containing $|{\bf v}|$ tokens, then $p_{i}$ is the probability that it belongs to player $i$. 
A probabilistic interpretation allows for the use of tools from various areas in statistics, including information theory \cite{cover2012elements}, for the purpose of analysing token distributions in games. We briefly introduce the entropy and relative entropy of a token distribution as metrics. 
\subsection{Entropy}

The Shannon entropy of a random variable $X$ distributed according to $p(i)$ is defined as 
\begin{equation}
H(X)=-\sum_{i=1}^{n} p(i) \log_{2} p(i).
\end{equation}
Here, $H(X)$ is the average measure of uncertainty in the random variable $X$ and computes on average the number of bits required to describe the random variable $X$ \cite{cover2012elements}. The entropy is bounded according to 
\begin{equation}
0\le H(X)\le \log_{2} n
\end{equation}
If a one player owns all the tokens, then $H(X)=0$ and the answer to who owns the token drawn from the bin is known in advance. On the other hand, if $H(X)=\log_{2}n$, then the tokens are uniformly distributed amongst the agents, i.e, $p_{i}=1/n$, and a randomly drawn token could belong to any of the the $n$ players with equal probability. 
$H(X)$ gives us an idea of the weighting of a distribution between these extremes. 

\subsection{Relative Entropy}

Another useful metric is the relative entropy which is the measure of the distance between two distributions. Given $X\sim p(i)$ and $X\sim q(i)$. The relative entropy is
\begin{equation}
D(p||q)=p(x)\sum_{i}\log_{2}\frac{p(i)}{q(i)}\cdot
\end{equation}
The relative entropy is useful in the following context:~Suppose that at round $r$ of a token exchange game $n$ players have the token distribution $X\sim q(x)$ and the game ends when players attain the distribution $X\sim p(x)$. On average exchanges between successive rounds reduce the relative entropy by $\ell$ bits. How many more rounds will the game last? If $k$ is the number of rounds required, then $k=\ell^{-1}D(p||q)$.  Further applications could be, for example, in a blockchain where a governance aim is to modulate the entropy of the token distribution. 

Information theory, is a rich subject in its own right and there are several more useful quantities from information theory that can be developed for studying  and interpreting token distributions \cite{cover2012elements}. However, this is best left for future work.


\section{Multilayer Games}
\label{mlg}

\subsection{Token Portfolio Vectors}

The usual arguments for price in economics are centred around supply and demand. The seller sets a supply curve and, concomitantly, the buyer sets a demand curve. At the intersection of the two curves the `price' of exchange is set in terms of monetary tokens. In a token exchange game this type of exchange rule is an agreement that takes place across multiple layers.
Consider the following example: Players $A$ and $B$ are both in the layers $\{\textsc{widget}, \textsc{doodad}\}$. In round $r$, $A$ agrees to give $a$ \textsc{widgets} to $B$ and $B$ agrees to give $b$ \textsc{doodads} to $A$. The respective exchanges happen vertically within the layers, but the transfer of ownership appears to happen horizontally across layers: There is a change in the `token portfolio' of the players -- for players that act across several layers the experience is of having a certain number of tokens of this type, and of that type, and so forth. The idea of a token portfolio can be formalized as follows: 

\begin{define}
	\label{tpv}
	Token portfolio vector. Suppose that a player $P$ holds (possibly zero) token balances in a set of layers $J_{P}\subseteq J$, where $J$ is the set of potentially available layers in round $r$ of a multilayer game. Let $|J_{P}|=k$ and set $J_{P}=\{s_{1},s_{2},\ldots,s_{k}\}$. Player $P$ has the token portfolio vector ${\bf z}_{rP}=(P_{s_{1}},P_{s_{2}},\ldots, P_{s_{k}})$ where $P_{s_{i}}$ is the player's balance in layer $s_{i}$. 
\end{define} 
Continuing our example, we may suppose that, initially, $\vec{z}_{rA}=(a,0)$ and $\vec{z}_{rB}=(0,b)$. Once the exchange of \textsc{widgets} and \textsc{doodads} has been recorded we have, by construction, $\vec{z}_{(r+1)A}=(0,b)$ and $\vec{ z}_{(r+1)B}=(a,0)$. 

\subsection{Exploring Fungibility}

In an economic sense, two objects are fungible if they can be interchanged in some ratio. In the example above that ratio is $a:b$. Thus, in a token exchange game that has multiple layers, a natural question is what an exchange rate means, and how this is implemented in the model. Each graph $G_{rs}\in G(r,s)$ is uniquely labelled. However, given two different values of $s$, for example, $s_{1}$ and $s_{2}$ the graphs $G_{rs_{1}}$ and $G_{rs_{2}}$ may have agents (represented by vertices) that are common. In which case, formally, if we assume that the agents are identifiable as their vertices, $V_{rs_{1}}\cap V_{rs_{2}}\ne \emptyset$. These shared agents make tokens in different layers fungible if there exists some prescribed ratio at which token exchanges between agents in layer $s_{1}$ are anti-correlated (exchange edges have different directions in each layer) with token exchanges in layer $s_{2}$. This motivates the next definition. 

\begin{define}
	\label{pw-f}
	Pairwise fungibility. A pair of token types $\{s_{1}, s_{2}\}$ are pairwise fungible if there is a finite rate $\rho_{s_{1}s_{2}}$ such that the exchange of a unit token in the $s_{1}$ layer induces an exchange of $\rho_{s_{1}s_{2}}$ tokens in the $s_{2}$ layer. The value $\rho_{s_{1}s_{2}}$ is the exchange or fungibility rate between the tokens, and the reciprocal rate $\rho_{s_{2}s_{1}}=\frac{1}{\rho_{s_{1}s_{2}}}$. Thus, if two ledgers are not fungible, then formally $\rho_{s_{1}s_{2}}=0$ and $\rho_{s_{2}s_{1}}=\infty$ or visa versa.
\end{define}
Note that this definition is an ideal and assumes that the fungibility rate holds for at least a single round and is averaged over participating players. Individual players may negotiate their own rates for a particular round and choice of token types, and information on the best available rates may not always be available to all players. There is also the possibility of trades that mix multiple token types, for example, two apples and a glass of milk for a cookie. We ignore this possibility for now and focus on the pairwise relationships between layers. The reciprocity of the fungibility rates is tantamount to the requirement that you can return something for the same price that you bought it. Finally, the notation above is suggestive: We may construct a fungibility matrix $\rho_{r}=[\rho_{ij}]_{r}$ which reflects the fungibility rates between multiple layers in round $r$. We use the conventions that $\rho_{ij}\ge 0$, and, by design, $\rho_{ii}=1$. The matrix $\rho$ now gives us a basic tool for modelling the meta-structure of multilayer games.

\subsection{Multilayer Games}

A multilayer game is a collection of single layer games that are indexed by a parameter. Recall from Definition~(\ref{teg}) that each member of the collection is indexed by $s\in {J}$ and, as with single layer games, there is an index for the round of the game $r\in { I}$. This corresponds to the two parameter family of graphs ${\mathcal G}(r,s)$ where $r\in {I}$ an $s\in { J}$. We suppose that each graph $G_{rs}\in {\mathcal G}(r,s)$ has a vertex set $V_{rs}$ and an edge set $E_{rs}$. Further, $|V_{rs}|=n_{rs}$ and $|E_{rs}|=m_{rs}$. As before, we may define weight functions on the vertices and edges which represent balances and transaction weights respectively.

As we have discussed in this section already, in a genuine multilayer game identical players may play across layers. Suppose that ${\mathcal P}(r,s)$ is a collection of sets of player labels in a multilayer game with $P_{rs}\in {\mathcal P}(rs)$ as the set of players that are active in round $r$ for layer $s$. In a game, given $G_{rs}\in {\mathcal G}(rs))$, each vertex $v\in V_{rs}$ must belong to some player $p\in P_{rs}$. Moreover, for layers $s$ and $s'$ to be fungible $P_{rs}\cap P_{rs'}\ne \emptyset$. This motivates the following:

\begin{define}\label{i-layer}
	Isolated layer. Suppose $G_{rs}\in {\mathcal G}(r,s)$ and ${\mathcal P}(r,s)$ is the set of player labels. If for all $r\in {I}$ for every $s'\ne s\in {J}$, we have $P_{rs}\cap P_{rs'}=\emptyset$, then $s$ is called an isolated layer in ${\mathcal G}$. 
\end{define}
An isolated layer is simply a layer that is not fungible with the rest of the game because the set of players in the layer do not overlap with other layers. Trivially, a layer may be empty of players because it is empty of tokens. The lack of a specific resource in board-games where the bank is an external treasury is a an example: 

\smallskip
\begin{center}
	\begin{minipage}[c]{0.8\linewidth}
		\textsf{A: Have you got any wheat?}\\
		\textsf{B: No, nothing. No wheat's been dealt yet.}
	\end{minipage}
\end{center}
\smallskip

\noindent The following result holds as a direct consequence of Definition~(\ref{i-layer}). 

\medskip
\noindent{\bf Lemma~A} \emph{If $s$ is an isolated layer, then for every choice of $s'$, the pair $\{s,s'\}$ are not fungible, i.e. $\rho_{ss'}=\infty$ or $0$.}

\smallskip
\noindent{\bf Proof of Lemma~A:}~Suppose that $s$ is an isolated layer. Then, $s$ shares no players with other layers (including players with zero balances). Since there are no common players between $s$ and any other layer $s'$ exchanges in $s$ cannot be correlated using inter-layer player rules with exchanges in any other layer $s'$. Thus, if $s$ is an isolated layer then for each choice of $s'$ the pair $\{s,s'\}$ has fungibility rate $\rho_{ss'}=\infty$ or $0$.~$\square$

\medskip
\noindent Lemma~A is the formal assertion that only players which are common to multiple layers and have something to exchange can create and implement between layer rules. We can put Lemma~A to use by defining the following graph: 

\begin{define}\label{fg}
	Fungibility graph family. Suppose that ${\mathcal G}(r,s)$ is a family of graphs in a multilayer token exchange game. A fungibility graph is a graph $H_{r}$ such that each vertex $v\in V(H_{r})$ corresponds one-to-one with an $s\in J$. Thus, the vertices in $H_{r}$ are the layers in ${\mathcal G}(r,s)$ in a given round $r$.  
	The edges of $H_{r}$ are weighted and directed depending on the pairwise fungibility relationships between layers labelled by $s\in J$ in round $r\in I$. In particular, if layer $s$ is not fungible with layer $s'$ during round $r$, then there is no edge from $s$ to $s'$ in $H_{r}$. Thus, the fungibility structure of a token exchange game with graph family ${\mathcal G}(r,s)$ can then be modelled using the fungibility graph family ${\mathcal H}(r)$.
\end{define}

In order to explore this concept further, consider the next example: Suppose that ${\mathcal G}(r,s)$ is family of graphs in a multilayer token exchange game with a fixed set of layers $S=\{s_{1},\ldots,s_{\ell}\}$. The layers are the vertices of each fungibility graph $H_{r}\in {\mathcal H}(r)$ and we have $S=V(H_{r})$ and $|V(H_{r})|=\ell$. If, in round $r$, every layer $s\in S$ is an isolated layer, then $H_{r}$ is an empty graph with $E(H_{r})=\emptyset$. On the contrary, if all layers are pairwise fungible, then $H_{r}$ has every possible edge, $\ell^{2}$ in total. For a specific example, let $S=\{s_{1},s_{2},s_{3}\}$ during round $r$. This allows us to construct a fungibility matrix $\rho_{r}=[\rho_{ij}]_{r}$ where each of $\rho_{ij}$ is the exchange ratio between layer $i$ and layer $j$. Suppose we have
\begin{equation}
\rho_{r}=\begin{bmatrix}
1& 10 & \infty\\
4&1 & 5\\
6& 7 & 1
\end{bmatrix}_{r}.
\end{equation}
\noindent The diagonal entries are necessarily equal to one, since individual layers are perfectly fungible with themselves. Now, consider the entry $\rho_{13}=\infty$. Based on our definitions this means that $s_{1}$ cannot be swapped for $s_{3}$: Players in $s_{1}$ will only accept an `infinite' amount of $s_{3}$ in order to give up units of $s_{1}$. Next, consider the entry $\rho_{31}$. This implies that players in $s_{3}$ would be willing to give up six units of $s_{3}$-tokens for each unit in $s_{1}$-tokens they might receive. If this were a marketplace conversation it would go something like: 

\smallskip
\begin{center}
	\begin{minipage}[c]{0.8\linewidth}
		{\sf A: I don't want any apples for my dollars.}\\
		{\sf B: I'll give you 6 apples for a dollar.}
	\end{minipage}
\end{center}
\smallskip

\noindent
The presence of many market players and variability in preferences allows for this situation to resolve itself. If no-one is willing to make this type of exchange and player $B$ will be forced to reduce his prices to a limit value of $0$. In which case, in round $r+1$, the matrix could read 
\begin{equation}
\rho_{r+1}=\begin{bmatrix}
1& 10 & \infty\\
4&1 & 5\\
10^{-3}& 7 & 1
\end{bmatrix}_{r+1}.
\end{equation}
On a broad scale, this the mechanisms of economics work toward establishing fungibility between different ledgers. The key purpose of a token exchange game is to make sense of this `informational organization'. Consider the entries $\rho_{12}=10$ and $\rho_{21}=4$. This implies that a players from layer $s_{1}$ are willing on average to exchange 1 unit of $s_{1}$-tokens for 10 units of $s_{2}$ tokens. On the other hand, players in $s_{2}$ are willing to exchange 4 units of $s_{2}$-tokens for a single $s_{1}$-token. Effectively, this is the following: 

\smallskip
\begin{center}
	\begin{minipage}[c]{0.95\linewidth}
		{\sf A: I'll put a dollar in your dollar account, if you put ten fish in my fish account.}\\
		{\sf B: I'll give you four fish in your fish account for a dollar in my dollar account.}
	\end{minipage}
\end{center}
\smallskip

\noindent
This discussion is akin to a bid-ask spread: In practice agents iteratively produce prices that meet somewhere in the middle, so as to optimize their utility or reduce their uncertainty. This motivates the notion of a `local fungibility equilibrium' where there is a reciprocal fungibility between layer pairs and $\rho_{ij}=1/\rho_{ji}$.
\begin{define}\label{lfe}
	Local fungibility equilibrium. The pair of layers $\{s, s'\}$ are said to be in a local fungibility equilibrium if and only if their fungibility rates $\rho_{ss'}$ and $\rho_{s's}$ are reciprocals:~$\rho_{s's}={1}/{\rho_{ss'}}$.
\end{define}

\subsection{Bargaining}
\label{bargain}

A grasp of fungibility in the abstract sense is not useful without a model for how fungibility ratios are set: At some level one would like to know how (and why) an edge in the fungibility graph from Definition~(\ref{fg}) is constructed. On the other hand, pricing is arguably an arbitrary process. We give examples of how prices can be set by players below.

\subsubsection{Auctions}

Suppose that ${\mathcal G}(r,s)$ is a family of graphs in a multilayer game and $r\in I$, with $s\in J$. Let $\mathcal{P}(r,s)$ be the family of player sets for the game. Let $x\in J$ and suppose that $x$ is the `auction item' layer. Let $J^{*}\subset J\backslash\{x\}$ such that for any layer $s^{*}\in J^{*}$ and a given graphs $G_{rs*}$ and $G_{rx}$ for each $s^{*}\in J^{*}$ we have $P_{rs^{*}}\cap P_{rx}\ne \emptyset$. Then, $J^{*}$ is a set of `bid layers' and fungibility is established as follows:

\begin{itemize}
	\item Immediately prior to round $r$ players in $x$ declare to an oracle that they are willing to accept minimum bids $\beta_{s^{*}}$ where $s^{*}\in J^{*}$ for a quantity of $\alpha$ items in layer $x$. 
	\item The oracle shares this information with players in $J^{*}$ who are allowable bidders. 
	\item Players in the bid layers make bids accordingly and the winning bid $\beta^{*}_{y}$ where $y\in J^{*}$ or no bid is declared to the oracle. 
	\item The oracle then declares that the items are fungible between layers $x$ and $y$ in a ratio of $\alpha:\beta_{y}^{*}$ for round $r$ implying that $\rho_{xy}=\beta_{y}^{*}/\alpha$. 
\end{itemize}
This type of auction is just one way of establishing the equivalence of two layers. Another may be taking the average of various individual bids; preselecting players depending on balances and so forth.

\subsubsection{Random Ratios}

Suppose that ${\mathcal G}(r,s)$ is a family of graphs in a multilayer game and $r\in I$, with $s\in J$. Let $\mathcal{P}(r,s)$ be the family of player sets for the game. Let $x,y\in J$. Further, suppose that $P_{rx}\cap P_{ry}\ne \emptyset$ for some $r$. Let $\phi_{rx}, \phi_{ry}$ be vertex weightings. For any $r\in I$ and $s\in J$, given a player $p\in P_{rs}$, let $\phi_{rs}(p)$ be the total weight of all vertices belonging to player $p$ in the graph $G_{rs}$. Consider the preselected player sets $A_{rx}\subset P_{rx}$ and $B_{rx}\subset P_{ry}$ and suppose that for $v\in A_{rx}$, $\phi_{rx}(v)\ge \kappa$, and for $w\in B_{rx}$, $\phi(w)\ge \alpha$. 
Next, suppose that a regulator imposes the following fungibility condition immediately before round $r$. 

\begin{itemize}
	\item Each player in $A_{rx}$ is given a $\kappa$ sided die.   
	\item Each player in $B_{ry}$ is given an $\alpha$ sided die. 
	\item The players roll their respective dice and the results are declared to an independent oracle. 
	\item The oracle uses the average of $|A_{rx}|$ player rolls  to produce a value $X_{A}$ such that $1\le X_{A} \le \kappa$ and, similarly, the average of $|B_{ry}|$ player rolls to produce a value $Y_{B}$ such that $1\le Y_{B}\le \alpha$.
	\item The oracle declares that the layers $x$ and $y$ are fungible for round $r$ at a ratio of $X_{A}:Y_{B}$ so that $\rho_{xy}=Y_{B}/X_{A}$ and  $\rho_{yx}=X_{A}/Y_{B}$. 
\end{itemize}
This method of setting fungibility rates simply requires a global player agreement, a trustworthy oracle and a trusted channel so that the results of each roll may be recorded correctly. 

\subsubsection{Blind Bargaining}

Here is another game with a different structure. Suppose that we have a multilayer game with family ${\mathcal G}(r,s)$ and player family $\mathcal{P}(r,s)$ for $r\in I$ with  $s\in J$. For $x,y\in J$ construct the non-empty player sets $A_{rx}\subset P_{rx}$ and $B_{ry}\subset P_{ry} $ such that $A_{rx}\cap B_{ry}=\emptyset$. Let $T_{x}$ and $T_{y}$ be the total tokens in circulation in layer $x$ and layer $y$ just before the start of round $r$. Suppose that a fungibility rate is determined as follows:

\begin{itemize}
	\item Each player in $A_{rx}\cup B_{ry}$ is given two numbers by an oracle, say $\alpha$ and $\beta$, where $0<\alpha<\beta$ and where $\alpha$ and $\beta$ are much smaller than $T_{x}$ and $T_{y}$. 
	\item The players choose either $\alpha$ or $\beta$ and submit it to the oracle in a vote. 
	\item The oracle determines a ratio $X_{A}:Y_{B}$ by counting the number of $\alpha$'s and $\beta$'s submitted by the players. The value $Y_{B}$ is determined by the players in $A_{rx}$ by majority vote, and the value $X_{A}$ is determined by the players in $B_{rx}$ by majority vote. The oracle declares the result and the fungibility rate is fixed for round $r$ as $\rho_{xy}=Y_{B}/X_{A}$ and $\rho_{yx}=X_{A}/Y_{B}$. 
\end{itemize}
The resulting ratios will either be $\alpha:\beta$, $\beta:\alpha$ or $1:1$ thus fixing the rate $\rho_{xy}$ as $\alpha/\beta$, $\beta/\alpha$ or $1$. 

\subsection{Fungibility Graphs}

In Section~\ref{bargain} several examples of the bargaining process are developed. In this process, multiple agents collaborate across layers and create \emph{local contracts} -- locally binding rules for the exchange of tokens across different layers of a game for a limited number of rounds. Further, various real-world agreements including debt instruments operate in this way with a rule existing for the duration of a contract.  

The question then, is how does one consider the fungibility structure between layers? Definition~(\ref{fg}) introduces the idea of a fungibility graph: Recall that given the family ${\mathcal G}(r,s)$ with $r\in I$ and $s\in J$ (and appropriate weightings) we can set up a fungibility graph family ${\mathcal H}(r)$ that tracks the networked exchange structure between the layers. In round $r$, the graph $H_{r}\in {\mathcal H}(r)$ has vertices drawn from the layers $s\in J$ and $V(H_{r})=J_{r}\subseteq J$ where $J_{r}$, is the set of layers that are active in round $r$. 
On the other hand, the edge structure of the graph $H_{r}$ requires a locally binding contract, as well as a a weighting for that contract. For example, if a rule exists between $i,j\in V(H_{r})$ such that layers $i$ and $j$ are fungible then $e_{ij}=ij\in E(H_{r})$ and $e_{ji}=ji\in E(H_{r})$. Moreover, the rule specifies two weights $\rho_{ij}$ and $\rho_{ji}$ which govern the exchange rates between the layers. Now, if the rates are reciprocals of each other, then $\rho_{ij}=1/\rho_{ji}$ and a local fungibility equilibrium exists. If this equilibrium exists for any pair of vertices $i,j\in V(H_{r})$, then the resulting graph is no longer directed. Further, if self-edges are fixed (layers are self-fungible in a 1:1 ratio) and we may ignore these edges, then the graph $H_{r}$ is a simple graph.  

Section~\ref{bargain} illustrates that establishing contracts for exchange requires the exchange of information between multiple parties. A natural step forward, when considering the fungibility graph $H_{r}$ and family ${\mathcal H}(r)$ is a weighting of edges that captures the \emph{information theoretic cost} of establishing a contract and is measured in bits. We suppose that there is some cost function $\kappa_{ij}$ associated with creating the edge $ij\in E(H_{r})$. We impose the restriction that fungibility (except for self-fungibility) cannot be established locally without a contract and without the exchange of information. In particular, if $\kappa_{ij}=0$ layers $i$ and $j$ are not fungible. In a similar vein to $\rho_{r}=[\rho_{ij}]_{r}$, the fungibility matrix for round $r$, we may also construct, $\kappa_{r}=[\kappa_{ij}]_{r}$ as a cost matrix for round $r$. Note that $\kappa_{ij}\ge 0$, $\kappa_{ii}=0$, and if the graph $H_{r}$ is a simple graph, then $\kappa_{ij}=\kappa_{ji}$ and the matrix is symmetric. Also, the more complex a contract, the higher the cost in terms of information exchange. Further, these assumptions imply that the adjacency structure of $H_{r}$ is unchanged whether we consider $\kappa_{r}$ or $\rho_{r}$. 

In a similar vein, we may introduce another cost measure $\mu_{ij}$, where $i,j\in V(H_{r})$, for establishing the fungibility between layers $i$ and $j$ as a measure the the information required to prevent an \emph{arbitrage} opportunity from being created in round $r$, i.e. an agent cannot create a contract in which a unit of a token in layer $x$, is exchanged for tokens in layer $y$, which are exchanged for tokens in layer $z$, which are in turn exchanged for more than one unit of tokens in layer $x$ all in the same round. Unlike, $\kappa_{ij}$, we may have $\mu_{ij}=0$. We may also construct the matrix $\mu_{r}=[\mu_{ij}]$ which is a weighting (measured in bits) on the graph $H_{r}$.   
Using these constructions allows us to prove the following result. 

\medskip
\noindent{\bf Theorem B.}~\emph{Given a token exchange game represented by the graphs $G_{rs}\in {\mathcal G}(r,s)$ and a concomitant fungibility (simple) graph $H_{r} \in {\mathcal H}(r)$ where $r\in I$ and $s\in J$. If $H_{r}$ is arbitrage minimising, then either $H_{r}$ is a tree or $\mu_{r}=[0]_{r}$ with $\mu_{ij}=0$ for any $i,j\in V(H_{r}).$}
\medskip

\noindent{\bf Proof of Theorem B.}~By assumption, $H_{r}$ is a simple graph. Suppose that $H_{r}$ is arbitrage minimizing. Without loss of generality, we may assume that $H_{r}$ is connected. If $H_{r}$ is a tree, then there are no cycles, and therefore no opportunities for arbitrage in round $r$ and so $\mu_{r}=[0]_{r}$. Thus, assume that $H_{r}$ is not a tree. In this case, $H_{r}$ contains a cycle $C$. Let $e=xy\in E(C)$ be an edge on $C$. Since $e$ is contained on a cycle, $e$ is not a bridge. Thus, $H_{r}-e$ is connected. Let $\Sigma(H_{r})=\sum_{e\in E(H_{r})} \mu(e)$, where $\mu(e)$ is the arbitrage cost of the edge $e$. Since $\mu(e)\ge 0$, we must have $\Sigma(H_{r}-e)\le \Sigma(H_{r})$. Removing edges from $H_{r}$ until a minimal spanning tree $T$ is obtained gives us a tree such that $V(T)=V(H_{r})$, and $\Sigma(H_{r})=\Sigma(T)$ or $\Sigma(T)< \Sigma(H_{r})$. The latter implies a contradiction. Thus, either $\Sigma(H_{r})=\Sigma(T)$ or $\mu_{r}=[0]_{r}$, as required.~$\square$

\medskip

Theorem~B implies that either the `informational cost' of establishing that there are no arbitrage opportunities within a given fungibility graph is zero or the graph $H_{r}$ is a acyclic, e.g. a tree or a forest.  If the informational cost of establishing that there are no arbitrage opportunities within the structure is  
zero, then this could be the effect of a global rule that affects multiple layers. For example, a system imposed penalty that reduces arbitrage opportunities to zero (assuming that this can be implemented effectively). On the other hand, if the graph $H_{r}$ is a tree or a forest, then $H_{r}$ contains no cycles and therefore no opportunities for arbitrage. It is, arguably, on this basis that single currency economies and monetary systems emerge historically as opposed to barter economies. 

\begin{figure}[t!]
	\label{fig00}
	\begin{center}
		\includegraphics[scale=0.7]{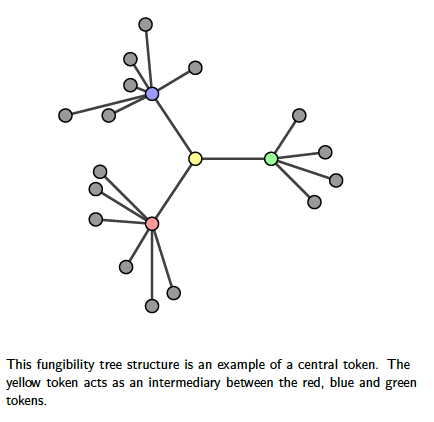}
	\end{center}
	\caption{Star Graph as a Central Clearing Token}
\end{figure}


\section{Velocity Metrics}
\label{mm}

The framework above is more useful if metrics can be developed for measuring standard economic quantities during a token exchange game. In this section, we use algebraic methods to consider various aspects of token velocity which has often been cited as a means to value cryptotoken based assets \cite{buterin,burniske}. 

\subsection{Using the Trace}
\label{mm1}
Suppose that we are given our usual family of graphs ${\mathcal G}(r,s)$. Recall that  in a game $G_{rs}\in {\mathcal G}(r,s)$ has the edge weighting matrix $W=[w_{ij}]_{rs}$ for the graph $G_{rs}$. A straightforward metric that we can use is related to the trace function on $W$. Suppose that there are $n_{rs}$ vertices in $G_{rs}$ in layer $s$ during round $r$ and that each vertex  in $v\in V_{rs}$ has positive weight with $\phi_{rs}(v)>0$. Setting
\begin{equation} \zeta_{rs}=\frac{\textrm{tr}(W)}{n_{rs}}=\frac{\sum_{i} w_{ii}}{n_{rs}},\label{zeta}\end{equation}
we have $0\le \zeta_{rs} \le 1$. This gives us a measure of `circulating' tokens in layer $s$ since it sums the diagonal entries via the trace: If $\zeta_{rs}=0$, then all tokens are circulating and if $\zeta_{rs}=1$, then no tokens are circulating. A complementary metric is $\zeta_{rs}^{*}=1-\zeta_{rs}$. The pair $\zeta_{rs}$ and $\zeta_{rs}^{*}$ measure, respectively, the proportion of tokens retained and circulated in a given round and layer. 

For a global metric, recall that $r\in I$ and $s\in J$ and the a game evolves through these index sets. At any point, we may interested in the level of circulation in a particular subset of rounds and layers $I^{*}\subseteq I$ or $J^{*}\subseteq J$. The metric $\zeta_{rs}$ may be generalised over several values of $r$ and $s$ as
\begin{equation}\zeta=\frac{\sum_{r\in I^{*}}\sum_{s\in J^{*}}\zeta_{rs}}{|I^{*}||J^{*}|}.\label{zeta2}\end{equation}
By ranging over several values of $r$ and $s$, we are investigating the matrix $M_{\zeta}=[\zeta_{rs}]$, where $r\in I^{*}$ and $s\in J^{*}$, and whose entries lie between zero and one. Further, $M_{\zeta}$ has dimensionality $|I^{*}|\cdot |J^{*}|$. The complementary matrix $M_{\zeta^{*}}$ may be constructed similarly. In most real world games, circulation volumes will be hidden, however on publicly available blockchain data, this metric can be directly computed over blocks (rounds) and token types (layers).  

Finally, note that the assumption that each vertex has positive weight is necessary to avoid skewing the metric, but can be circumvented by using the partial trace over a subgraph of $G_{rs}$: In some games token holders with a balance of zero that reassign a zero balance to themselves need to be excluded from the computation.  This can be done using taking partial trace over those diagonal elements for vertices $v\in V_{rs}$ for which $\phi_{rs}(v)>0$ and using the subgraph $K_{rs}\subset G_{rs}$ for which this condition holds. The value of $n_{rs}$ in Eq.~(\ref{zeta}) above is replaced by $|V(K_{rs})|$. This method still takes into account transfers of tokens to players with zero balances, but it does not skew the metric in favour of token holding.

\subsection{Inflation and Deflation}
\label{infdefsec}

In all economies the experience of inflation is that of increased prices. There are two sources which are commonly responsible: Increases in the money supply or a reduction of productivity. In the former, if the money supply increases then, other things being equal, there are more tokens chasing the same amount of production and the ratio of tokens to goods has changed. In the latter, there are fewer goods and services, the same amount of tokens and the ratio of goods and services to tokens changes. Thus, in order to explain inflation using our framework, it will be necessary to look at the fungibility of a pair of ledgers in tandem and consider their fungibility ratio as the supply of tokens in each ledger is allowed to vary. 

Suppose that we have a token exchange game with token type set $J=\{\textsc{widgets}$, $\textsc{doodads}\}$ and where the game runs over rounds $I=\{0, 1,\ldots,k\}$.  This induces the family of graphs ${\mathcal G}(r,s)$ where $r\in I$ and $s\in J$. For brevity, we may suppose that $s_{1}=\textsc{`widgets'}$ and $s_{2}=\textsc{`doodads'}$ so $J=\{s_{1},s_{2}\}$. 
In terms of fungibility, we are interested in the rates at which \textsc{widgets} are exchanged for \textsc{doodads} and visa versa. Using the language of the previous section, we may define a series of $2\times 2$ matrices $\rho_{r}$ where $r\in I$ to describe the fungibility of the different token layers. 
If the pair is uncoupled, and there is no exchange price for some round $r$, then 
\begin{equation}
\rho_{r}=\begin{bmatrix}
1& \infty \\
0&1 \\

\end{bmatrix}=\begin{bmatrix}
1&0\\
\infty & 1
\end{bmatrix}.\label{rho}
\end{equation}
The diagonal entries are `self' exchange rates, while the off diagonal entries are exchange rates between the different ledgers. We will rule out this exceptional case, and suppose that a finite fungibility rate can always be established between our ledgers. 
Thus, we are interested in the behaviour of the matrices $(\rho_{1}, \rho_{2}, \ldots, \rho_{k})$ and, in particular, in the behaviour of the generic fungibility matrix
\begin{equation}
\rho_{r}=\begin{bmatrix}
1& x_{r} \\
\frac{1}{x_{r}}&1\end{bmatrix}
\end{equation}
where, by convention, we set $x_{r}$ to be the \textsc{widgets} that can be bought per \textsc{doodad} in round $r$, and so $1/x_{r}$ is amount of \textsc{doodads} that can be bought per \textsc{widget}.

The next question is how the value of $x_{r}$ is calculated from the aggregated buyer-seller decisions that are made across the individual layers. Suppose that \textsc{widgets} are only swapped for \textsc{doodads} and \textsc{doodads} for \textsc{widgets} with no other drivers for changes in allocations other than mutual exchanges between respective token holders. Let $\chi_{rs_{1}}$ be the total number of circulating \textsc{widgets} and suppose $\chi_{rs_{2}}$ is the total number of circulating \textsc{doodads} in round $r$. Let $\zeta_{rs_{1}}$ and $\zeta_{rs_{2}}$ be the circulation metrics defined in Section~(\ref{mm1}) for their respective token sets. 
In a given round, the following quantities give us the tokens that are unsold in their respective ledgers:
\[\zeta_{rs_{1}}\chi_{rs_{1}}\qquad\textrm{and}\qquad \zeta_{rs_{2}}\chi_{rs_{2}}.\] 
On the other hand, the values 
\[(1-\zeta_{rs_{1}})\chi_{rs_{1}}\qquad\textrm{and}\qquad (1-\zeta_{rs_{2}})\chi_{rs_{2}},\] represent that volume of tokens that are sold or exchanged in their respective ledgers. The fungibility rate emerges from the equivalence:
\begin{equation}(1-\zeta_{rs_{1}})\chi_{rs_{1}}\,\textsc{widgets}\equiv (1-\zeta_{rs_{2}})\chi_{rs_{2}}\,\textsc{doodads}.\label{wd1}\end{equation}
Dividing across gives values for $x_{r}$ and $1/x_{r}$ in the matrix for $\rho_{r}$. 

Since we are interested in price increases, we may work with the value of $x_{r}$ directly and consider the effect of changes in relative supply and demand of tokens in each ledger. We have 
\begin{equation}
x_{r}=\frac{(1-\zeta_{rs_{1}})\chi_{rs_{1}}}{(1-\zeta_{rs_{2}})\chi_{rs_{2}}}\cdot\label{infdef}
\end{equation} This is obtained from Eq.~(\ref{wd1}) by dividing through so that $x_{r}$ \textsc{widgets} is equivalent to 1 \textsc{doodad}. Now, it is relatively easy to track changes in price that are related to changes in supply: If the supply of tokens in the \textsc{widget} layer $\chi_{s_{1}}$ increases, then, all other things being equal, $x_{r}$ will increase. Similarly, if the value of $(1-\zeta_{rs_{1}})$, which captures the demand for \textsc{doodads} by \textsc{widget} holders would increase, the number of \textsc{widgets} per \textsc{doodad} will also increase. Alternatively, if the supply, $(1-\zeta_{rs_{1}})\chi_{rs_{1}}$, would decrease and the value $(1-\zeta_{rs_{2}})\chi_{rs_{2}}$ would increase, then $x_{r}$ would decrease since there are fewer \textsc{widgets} available per \textsc{doodad}. In turn, taking the reciprocal equation and considering changes in the value of $1/x_{r}$ over successive rounds gives an indication of the availability of \textsc{doodads} per \textsc{widget}. Note that the extreme cases in which there is no supply or no demand yield $x_{r}=\infty$ or $x_{r}=0$ as required by previous discussions on fungibility. 

\subsection{Token Velocity}

Most discussions around token velocity in the blockchain community  \cite{buterin,burniske} center around the equation of exchange in monetary economics which says that for a given period:
\begin{equation}
M\times V= P\times Q,
\end{equation}
where $M$ is the nominal money supply, $V$ is the velocity of money, $P$ is the price level and $Q$ is an index of `real' expenditures. The underlying premise is that, other things being equal, changes in the money supply has an impact on nominal price levels. In order to adapt this to a token exchange game, consider Eq.~(\ref{infdef}): Set $P=x_{r}$ and $M=\chi_{rs_{1}}$ with $V=(1-\zeta_{r_{s_{1}}})$ and $Q=(1-\zeta_{rs_{2}})\chi_{rs_{2}}$. So, Eq.~(\ref{infdef}) is the equation of exchange over a single round, and, further, it is invertible in the sense that it can be constructed from the $\textsc{widget}$ or the $\textsc{doodad}$ perspective. In principle, an equivalent equation for multiple rounds can be constructed using averages. 


\section{Examples}

\label{ex}

The most direct and interesting applications for the language developed thus far is in the area of cryptotokens and blockchain related products. The spectacular variety of cryptotoken products and the lack of clarity in terms of their behaviour makes a framework for studying their properties extremely valuable. In this section we apply the models developed above to study the Lightning Network \cite{LNW:2016} and the Circles cryptotoken system \cite{CRC}. 

\subsection{The Lightning Network}

The Lightning Network is a proposed upgrade to the original Bitcoin network that allows for scalability \cite{LNW:2016}. The idea is to create side chains that allow users to transact amongst themselves without necessarily recording all of their transactions to the main blockchain. 
The process works as follows: Two users (or more) within an existing ledger create multi-party channel and commit some of the their existing tokens to this channel. The value of the channel in the original ledger is the sum of the committed tokens. The channel then becomes a separate ledger and users within this ledger can then continue exchanging tokens without needing to load the main ledger with the record of these exchanges. Once the users are done with these side exchanges they exit the channel and final balances are uploaded to the main blockchain (the original ledger) and the channel is destroyed.

In order to capture this as a token exchange game, suppose that we have a family of graphs ${\mathcal G}(r)$ which represents a single layer token exchange game - the original ledger. Further, let the total number of nodes in round $r$ be $n_{r}$ and, without loss of generality, assume that the token supply is fixed at $T_{\mathcal G}$. 

Suppose that in round $r$ a total of $m_{r}< n_{r}$ users  wish to use the lightning network in order to transact. In order to do this a new ledger with $m_{r}$ nodes needs to be created and a node is added to the $G_{r}$ network so that $|V(G_{r+1})|=n_{r}+1$. The sum of the committed tokens is the balance reflected in $G_{r+1}$ for the new node. The total number of tokens $T_{\mathcal G}$ remains unchanged. 

A sublayer is created as a new game corresponding to this additional node and each of the $m_{r}$ users owns their committed tokens in the sublayer. Suppose that the sublayer corresponds to a family of graphs ${\mathcal F}(j)$ where $j$ indexes the rounds in the new layer. In any given round a vector ${\bf f}_{j}$ tracks the ownership of tokens in the layer. The initial token distribution vector in this layer ${\bf f_{0}}$ assigns the tokens committed in round $r$ of ${\mathcal G}(r)$ to agents in round $j=0$ of the new layer. Now the sublayer is out of sync with the original one and transactions take place in the sub layer. The total number of tokens in the sublayer are $T_{\mathcal F}\le T_{\mathcal G}$. Suppose that for the moment $T_{\mathcal F}$ is fixed and the game is closed (the open case is relatively easy to handle). Depending on the rules of the sublayer various matrices will track the exchanges in the layer and we may suppose that the relationship for a collection of (possibly stochastic) matrices $W_{j}$. 
\begin{equation}
{\bf f}_{j+1}=W_{j}{\bf f}_{j}
\end{equation}
We may suppose that the sublayer exists only for a finite number of rounds $j\in \{0,1,\ldots, k\}$ and the final distribution 
\begin{equation}
{\bf f}_{k}=W_{k-1}W_{k}\ldots W_{1}{\bf f}_{0}
\end{equation}
is uploaded to the original layer by adding the vector ${\bf f}_{k}$ (with zero filling for non-participants in the sublayer) to the distribution vector in round $G_{r+2}$ of the main layer. The sublayer place holding node is then deleted from the layer. 

The main idea here is nestedness and the operation of creating a sublayer from an existing layer. Rounds can occur within the sublayer with a distribution update to the main layer happening only later. The token supply to the sublayer can be varied as new funds are committed to the `parent' node in the main layer, and the dimensionality of the sublayer can be varied as new members join the family ${\mathcal F}(r,s)$ from the main layer. As long as $m_{r} \ll n_{r}$ this can be an efficient process. A detailed schematic is presented in Figure~2. 

\begin{figure}[t!]
	\label{fig1}
	\begin{center}
		\includegraphics[scale=0.625]{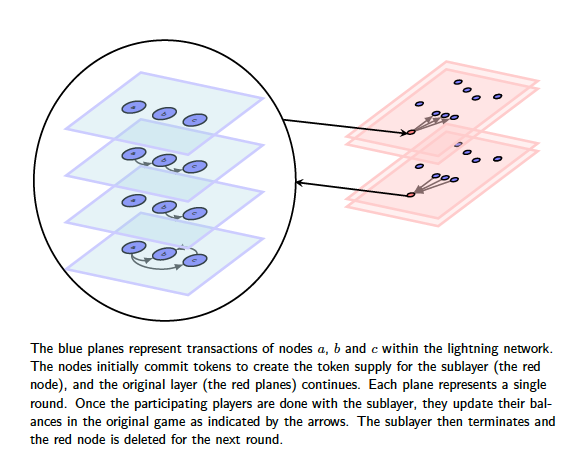}
	\end{center}
	\caption{The Lightning Network as a Token Exchange Game}
\end{figure}

\subsection{Circles}

Circles is a proposed blockchain based Universal Basic Income system \cite{CRC}. The idea is centred around personalized currencies and economic trust invested into each personal currency is due to underlying social trust. Every user is issued their own unique personal coin, and the system issues personal coins at a rate defined by an issuance algorithm. All coins are fungible at a fixed 1:1 ratio, but the question of whether someone will accept your coins and exchange them depends on social trust: The user is incentivized to hold tokens that are `trusted' by the underlying social trust network, and to establish trust into their own personal token. The value of a given token correlates to a users social credibility. 

As a general adaptation of circles into a token exchange game suppose that there are~$N_{r}$ users~in the circles community in a given round $r$ and so $N_{0}$ unique tokens at round $r=0$. This implies a multilayer game with $N_{r}$ layers in round $r$. Layers can be open or closed depending on whether there is an internal treasury (an extra player that acts as a coin repository) or an external treasury that issues the coins to the layers. In the following, we suppose that the game is open, and the tokens are issued from an external source. 
Then, the game operates according to the following rule set:
\begin{itemize}
	\item Issue a token to each layer in each round to the player that owns the layer from an external treasury. 
	\item Tokens are fungible between layers $i$ and $j$ during round $r$ in a fixed 1:1 ratio if token holders in layer $i$ trust token holders in layer $j$ based on an underlying social trust graph between all round $r$ players. Here, trust means that players are willing to exchange their respective personal coins for another's personal coins.
	\item No exchanges happen in round $r=0$.  
\end{itemize}

\noindent Recall that there are $N_{r}$ players in total in round $r$. In the network picture, this game induces the family of graphs ${\mathcal G}(r,s)$ where $r\in I$ and $s\in J$, and $I$ and $J$ are index sets. 
Further, $|J|\ge N_{r}$ since there will be $N_{r}$ unique circles tokens each with their own layer.  
Since there are $N_{r}$ players altogether, each graph $G_{rs}\in {\mathcal G}(r,s)$ has $|V(G_{rs})|\le N_{r}$. The decision about whose circles tokens to hold and when resides with the player and affects $|V(G_{rs})|$. For example, if in round $r$ every player decides to hold tokens Alice's layer $a\in J$, then $|V(G_{ra})|=N_{r}$. On the other hand, if no-one except Bob trusts tokens from Bob's layer $b\in J$ in round $r$, then $|V(G_{rb})|=1$. Thus, the graph structure, vertex weightings and edge weightings will depend on the fungibility structure between the layers. 

Suppose that ${\mathcal H}(r)$ is the family of fungibility graphs corresponding to ${\mathcal G}(r,s)$. We may suppose that $H_{0}$ is given based on some initial player relationships. In any given round $|V(H_{r})|=N_{r}$. We are interested in developing a model for how $H_{r}$ evolves for a general $r$. As the fungibility ratio is already fixed, we may simplify by ignoring the edge-weightings (fungibility rates) and focus solely on edge existence. 

A useful starting point is the model for preferential attachment in random graphs which is developed in the article by Barab\'asi and Albert \cite{BA:1999} and has been used extensively to model social networks. The model works as follows:
Given $H_{0}$ as an initial connected seed graph, we suppose that new nodes are added to the network at a rate of one per round. Each new node is connected to $n_{r}\le N_{r}$ existing nodes with a probability that depends on the degree of an existing node, i.e. people who know more people are more likely to know new people who join the network. The probability $p(v_{i})$ that a new node is connected to given node $v_{i}\in V(H_{r})$ is given by 
\begin{equation}
p(v_{i})=\frac{\delta(v_{i})}{\sum_{j}\delta(v_{j})}
\end{equation}
where $\delta(v_{i})$ is the degree of vertex $i$. The effect is that after $r+1$ rounds, the magnitude of the vertex set is $|V(H_{r+1})|=|V(H_{1})|+r=\bar{n}$, and the degree distribution -- the fraction of nodes of degree $k$ is given by a polynomial $P(k)\sim k^{-3}$. This gives an insight into the underlying trust structure as the network evolves. Further details on random graphs may be found in \cite{Bollobas:2009}. 

As a second point of analysis, given a circles game, we may also consider the construction of a bipartite simple graph which maps out who owns which tokens. Suppose again that we have $n_{r}$ players that are circles token holders in round $r$. This motivates the construction of a bipartite graph $F_{r}$. Partition the graph into sets $X$ and $Y$ such that $V(F_{r})=X_{r}\cup Y_{r}$. Let $X_{r}$ be the set of players, and $Y_{r}$ be the set of tokens in round $r$. We construct the edge set as follows: For vertices $x\in X$ and $y\in Y$ the edge $e=xy\in E(F_{r})$ if and only if player $y$ is a holder of token $x$. We can weight the edges using the player's current balance of a particular tokens to deepen the analysis. This ties up with the idea of a token portfolio vector from Definition~\ref{tpv}. In this case, the sum of the edge-weights of a vertex $x\in X$ is the total number of circles tokens owned by player $x$, while the sum of the edge-weights of a vertex $y\in Y$ is the total number of circles tokens of type-$y$ that have been issued. 

\begin{figure}[t]
	\label{circles}
	\begin{center}
		\includegraphics[width=8cm]{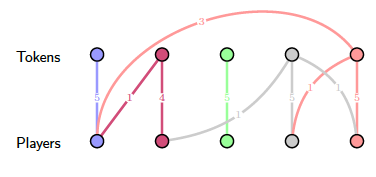}
	\end{center}
	\caption{A Circles Game}
\end{figure}

Consider the following example which has been adapted from~\cite{CRC}: The game in Figure~\ref{circles} has been going on for 9 rounds. Every minute a new personal coin issued. Eve (red) started the game at round 0, Diana (gray) joined in round 2, while Alice (blue), Bob (purple) and Charlie (green) joined in round 4. Alice has three {\sc evecoins} and one {\sc bobcoin} as well as five {\sc alicecoins}. In this representation, using the properties of random bipartite graphs such as degree, average degree, connectedness, and the plethora of other graph invariants  \cite{West:2017,Wilson:2015,Bollobas:2009,bornholdt2006handbook} we may deepen our understanding of the game as it evolves.

\section{Conclusion}

Token systems, at their heart, are systems in which records are exchanged amongst agents. The core result of this paper has been to develop a language for modelling these types of exchanges using networks and linear algebra. This provides a basis for understanding real world token systems including monetary and blockchain based token systems and their properties as mathematical entities through various metrics. This abstract understanding, in turn, has implications for the design, construction and analysis of token systems in the real-world.


\bibliography{research}
\bibliographystyle{ieeetr}
\end{document}